\documentclass[12pt,preprint]{aastex}
\usepackage{latexsym,psfig}
\newcommand{\etal}{et al. }
\newcommand{\teff}{{\mathrm{T}}_{\mathrm{eff}}}
\newcommand{\zzc}{ZZ~Ceti }
\received{2004 March 24}
\begin{document}

\title{Re-defining the Empirical \zzc Instability Strip}

\author{Anjum S. Mukadam, D. E. Winget, Ted von Hippel, M. H. Montgomery}
\affil{Department of Astronomy, University of Texas at Austin, Austin, TX -78712, U.S.A.; anjum@astro.as.utexas.edu}
\affil{McDonald Observatory, Fort Davis, TX 79734, U.S.A.}
\author{S. O. Kepler, A. F. M. Costa}
\affil{Instituto de F\'{\i}sica, Universidade Federal do Rio Grande do Sul, 91501-970 Porto Alegre, RS - Brazil.}

\begin{abstract}

	We use the new \zzc stars (hydrogen atmosphere white dwarf variables; DAVs) discovered within
the Sloan Digital Sky Survey (Mukadam \etal 2004) to re-define the empirical \zzc instability strip.
This is the first time since the discovery of white dwarf variables
in 1968 that we have a homogeneous set of spectra acquired using
the same instrument on the same telescope, and with consistent data reductions, for a
statistically significant sample of \zzc stars. The homogeneity of the spectra reduces the scatter
in the spectroscopic temperatures and we find a narrow instability strip of width
$\sim $~950\,K, from 10850--11800\,K.
We question the purity of the DAV instability strip as we find several non-variables within.
We present our best fit for the red edge and our constraint for the blue edge of the instability strip, determined
using a statistical approach.
\end{abstract}


\section{Introduction}
Global pulsations in white dwarf stars provide the only current systematic way 
to study their interiors. Hydrogen atmosphere white dwarfs (DAs) exhibit nonradial g-mode pulsations,
and are known as DA Variables (DAVs) or \zzc stars.
Bergeron \etal (1995, 2004) and Koester \& Allard (2000) find these pulsators confined
in the range 11\,000\,K and 12\,500\,K for
$\log g\approx 8$.

During the course of a 15 month long search, Mukadam \etal (2004, hereafter Paper-I) discovered 35 new
\zzc stars within the Sloan Digital Sky Survey (SDSS).
This is the first time in the history of
white dwarf asteroseismology that we have a statistically significant homogeneous set
of \zzc spectra, acquired entirely with the same detection system, namely the
SDSS spectrograph on the 2.5\,m telescope at Apache Point Observatory. All the
spectra have been reduced and analyzed consistently using the same set of model atmospheres
and fitting algorithms, including the observed photometric colors (see Kleinman \etal 2004).
This homogeneity should reduce the relative scatter
of the variables in the $\teff$--$\log ~g$ plane, and possibly allow us to see emerging new features.
The sample size of known DAVs is now almost twice as large since the last characterization of
the instability strip by Bergeron \etal (2004). 
However, we will not include the previously known DAVs in our
analysis with the exception of G\,238-53, as these pulsators do not have SDSS
spectra and will only serve to reduce the homogeneity of our sample.

We list the $\teff$ and $\log ~g$ values of all the variables and non-variables
we discovered within the SDSS data in Paper-I, along with their internal uncertainties.
Note that we will not be considering WD2350$-$0054 in this paper as it may be 
a unique pulsator; it shows pulsation periods and
pulse shapes characteristic of the hot DAV stars, while the SDSS temperature
determination places it below the cool edge
of the instability strip. We focus on the general
trends of the majority of the DAVs, and hence a discussion of
WD2350$-$0054 is postponed to a future date. We will not be including WD1443$-$0054
either, as its temperature and $\log ~g$ determinations are unreliable
due to a missing portion in its SDSS spectrum.
We will be including G\,238-53, the only previously known \zzc star with a published SDSS spectrum.

\section{Empirical instability strip} 
We show the empirical SDSS instability strip in Figure 1, as determined by 30 new \zzc stars and G\,238-53.
We plot histograms of the observed
variables as a function of temperature and $\log ~g$, and weighted histograms (see section 2.2) for the 
non-variables (Not Observed to Vary; NOVs). 
Figure 1 has two striking features: a narrow strip of width 950\,K and
non-variable DA white dwarfs within the
instability strip. 

Pulsations
are believed to be an
evolutionary effect in otherwise normal white dwarfs (Robinson 1979; Fontaine \etal 1985;
Fontaine \etal 2003; Bergeron \etal 2004).
Non-variables in the middle of the strip question this semi-empirical premise, even if we use
the uncertainties in temperature to justify the non-variables close to the edges.

We also note that the DAV distribution appears to be non-uniform across the strip.
The features of this plot
are influenced by various
factors such as biases in candidate selection, non-uniform 
detection efficiency in the $\teff$--$\log ~g$ plane,
and uncertainties as well as systematic effects in spectroscopic temperature and
$\log ~g$ determinations. We address these issues and their effects on the DAV
distribution in the next few
sub-sections.

\subsection{Biases in Candidate Selection}
We selected SDSS DAV candidates for high-speed photometry from those spectroscopically identified
DA white dwarfs that lie in the temperature range 11000--12500\,K. These temperature fits
are derived by our SDSS collaborators using the spectral fitting technique outlined in Kleinman \etal (2004). 
Paper-I gives
a discussion of other candidate selection methods used in our search for \zzc stars prior to the spectral fitting
technique. 

Our various science goals
lead to some biases in selecting DAV candidates for observation.
The hot DAV (hDAV) stars exhibit
extreme amplitude and frequency stability (e.g. Kepler \etal 2000a; Mukadam \etal 2003).
We plan to search for
reflex motion caused by orbiting planets around such stable pulsators
(e.g. Kepler \etal 1991; Mukadam, Winget, \& Kepler 2001; Winget \etal 2003). These stable clocks
drift at their cooling rate; measuring the drift rate in the absence of orbital companions
allows us to calibrate our evolutionary models. These models are useful in determining
ages of the Galactic disk and halo using white dwarfs as chronometers
(e.g. Winget \etal 1987; Hansen \etal 2002).
Therefore, we preferentially
choose to observe hDAV candidates in the range 11700--12300\,K to increase the sample of
known stable pulsators with both the above objectives in mind.
This bias is partially compensated for, as hDAVs are harder to find (see section 2.2).

We also preferentially observe DAV candidates of extreme masses.
Low mass ($\log g \leq 7.6$) DAVs could well be helium core white dwarfs; pulsating
He core white dwarfs should allow us to probe their equation of state.
High mass ($\log g \geq 8.5$) DAVs are potentially
crystallized (Winget et al. 1997; Montgomery \& Winget 1999), providing a
test of the theory of crystallization in stellar plasma.
Metcalfe, Montgomery, \& Kanaan (2004) present strong evidence that the massive DAV,
BPM\,37093, is 90\% crystallized.

The distribution of {\it chosen} DAV candidates also depends
on the distribution of {\it available} DAV candidates. We have an additional
bias due to the SDSS criteria in choosing candidates for spectroscopy.
But a histogram of the available DAV candidates
is consistent with a random distribution and does not reflect any systematic effects.

The non-uniform nature of the DAV distribution does not appear to be a candidate selection effect.
However, we are in the domain of small number
statistics since we observed only four DAV candidates in the range 11350--11500\,K.
Of these, two are massive and consequently expected to be low amplitude pulsators (see section 2.2),
making detection difficult.
Our data are suggestive of a bimodal DAV distribution in temperature.
We hope to investigate this issue further by observing additional DAV candidates in the range 11350--11500\,K
with our collaborators.

\subsection{Non-Uniform Detection Efficiency}
The hDAVs show
relatively few pulsation modes, with low amplitudes ($\sim $0.1--3\%) and
periods around 100--300\,s.
The cooler DAVs (cDAVs) typically show longer periods, around 600--1000\,s, larger amplitudes (up to 30\%),
and greater amplitude variability (Kleinman \etal 1998).
Massive pulsators show low amplitudes as a result of their high gravity ($\log ~g \geq 8.6$).
These distinct trends
of the pulsation periods and amplitudes with temperature and $\log ~g$
imply that the detection efficiency must also be a function of $\teff$ and
$\log ~g$. The detection efficiency not only varies in the $\teff$--$\log ~g$
plane, but is also dependent upon weather conditions and the magnitude of the DAV candidate.
Furthermore, a \zzc star may have closely spaced modes or multiplet structure, both of which cause beating effects.
Some of the non-variables in the instability strip could well be pulsators, that were in the low amplitude
phase of their beating cycle during the observing run.
McGraw (1977) claimed BPM\,37093 to be non-variable, but Kanaan et al. (1992)
showed that it is a low amplitude variable with evident beating. Dolez, Vauclair, \& Koester (1991)
claimed the non-variability limit of G\,30-20 to be a few mmag\footnote{One milli-magnitude (mmag)
equals 0.1086\% change in intensity.}, but
Mukadam \etal (2002) found G\,30-20 to be a beating variable with an amplitude
of 13.8\,mma\footnote{One milli modulation amplitude (mma) corresponds to 0.1\% change in intensity.}.

        In order to address these issues systematically, we simulate light curves of real pulsators
for different conditions and compute the resulting Fourier Transform (FT) to see if the signal 
is detectable above noise. We utilize the real periods and amplitudes,
with randomly chosen phases (to sample the beat period), to simulate two hour long light
curves\footnote{We
generally observe the DAV candidates for two hours at a time when searching for new variables.}.
We independently shuffle the magnitudes
and average seeing conditions of real data on the DAVs.
This ensures a realistic distribution for both these parameters.
We randomly select a magnitude and seeing value from these distributions
to simulate white noise, the amplitude of which is determined using a noise table based on real data.
We compute a FT of the light
curve and check if the star can be identified as a pulsator or if the signal was swamped by noise.
We repeat this procedure 100 times for each DAV for different phases, magnitudes, and seeing values.
Note that our noise simulation is not completely realistic, as it does not include effects due to variable
seeing, gaps in the data due to clouds, and low frequency atmospheric noise.
However, it does help us understand how the detection efficiency changes in the
$\teff$--$\log ~g$ plane.

We find that we are able to
{\it rediscover} almost all of the average and low mass cDAVs in the hundred simulated attempts.
The high mass ($\log ~g \geq 8.6$)
DAVs with a substantially
lower amplitude are recovered with a $\sim $70\% success rate. This implies that
non-variables in Figure 1 in the region
$\log ~g \geq 8.6$ have a 30\% chance of being low amplitude variables.
At the hot end of the instability strip,
both low pulsation amplitude and beating can cause us to miss even the
average or low mass hDAVs 35 out of 100
times.

Table 1 lists the
non-variables in the instability strip along with their temperature, $\log ~g$, magnitude,
and number of observing runs. The number after the NOV designation indicates
the best non-variability limit in mma. Based on the simulations, we assign each non-variable
a weight based on our estimate of the probability that the observed candidate is a genuine non-variable,
and not a low-amplitude or beating pulsator.
We use the non-variability limits to assign the weights 0.98, 0.95, 0.90, 0.85, 0.80, 0.70, and 0.60,
for NOV1, NOV2, NOV3, NOV4, NOV5, NOV6, and NOV7 or higher, respectively.
If the NOV is massive ($\log ~g \geq $8.6), then we additionally multiply its weight by a factor of 0.7.
If the NOV is close to the blue edge of the strip, then we multiply by a factor of 0.65 to
account for low amplitude and/or beating pulsators.
However if the NOV has been observed multiple times, then it is unlikely to have been missed
as a result of beating. In such a case, we multiply its weight
only by a factor of 0.8 instead of 0.65, to allow for a possible low amplitude variable.
We utilize these weights in section 6 to compute best-fit red and blue edges.

\clearpage
\begin{deluxetable}{ccccccc}
\rotate
\tablecolumns{7}
\tablewidth{0pc}
\tablecaption{Non-variables in the \zzc instability strip}
\tablehead{
\colhead{Object} & \colhead{Limit} & \colhead{Obs. Runs} &\colhead{SDSS $\teff$ (K)} & \colhead{SDSS $\log ~g$} &\colhead{$g$}  & \colhead{Weight} }
\startdata
WD0037+0031 & NOV5 & 2 &$10960 \pm 050$ & $8.41 \pm 0.03$ & 17.5 &  0.80 \\
WD0050$-$0023 & NOV6 & 2 &$11490 \pm 090$ & $8.98 \pm 0.03$ & 18.8 &  0.50\\
WD0222$-$0100 & NOV3 & 4 &$12060 \pm 120$ & $8.12 \pm 0.05$ & 18.0 & 0.60 \\
WD0303$-$0808 & NOV4 & 2 &$11400 \pm 110$ & $8.49\pm 0.06$ & 18.8 &  0.85 \\
WD0345$-$0036 & NOV5 & 3 &$11430 \pm 150$ &$ 7.76 \pm 0.09$& 19.0 &  0.80\\
WD0747+2503 & NOV3 & 3 & $11050 \pm 110$&$7.93 \pm 0.08$ & 18.4 &  0.90\\
WD0853+0005 & NOV4 & 2 & $11750\pm 110$ &$ 8.11\pm 0.06$ & 18.2 &  0.55\\
WD1031+6122 & NOV4 & 2 & $  11480 \pm 180$ & $7.68 \pm 0.11$ & 18.7 &  0.85\\
WD1136$-$0136 & NOV2 & 1 & $ 11710 \pm 070 $ & $ 7.96 \pm 0.04 $ & 17.8  & 0.62\\
WD1337+0104 & NOV4 & 2 & $ 11830 \pm 210$ &$ 8.39 \pm 0.11$ & 18.6 &  0.60\\
WD1338$-$0023 & NOV4 & 1 & $ 11650 \pm 090$ &$ 8.08 \pm 0.05$ & 17.1 &  0.85\\
WD1342$-$0159 & NOV4 & 2 & $ 11320 \pm 160$ &$ 8.42 \pm 0.09$& 18.8 & 0.85 \\
WD1345+0328 & NOV6 & 1 &$ 11620\pm 140$ &$ 7.80\pm 0.08$ & 18.6 &  0.70\\
WD1432+0146 & NOV5 & 1 &$11290 \pm 070$ &$ 8.23\pm 0.06$& 17.5 & 0.80 \\
WD1443$-$0006 & NOV5 & 1 & $11960\pm 150$ &$ 7.87\pm 0.07$ & 18.7 & 0.80 \\
WD1503$-$0052 & NOV4 & 3 & $ 11600 \pm 130 $ & $ 8.21 \pm 0.07 $&18.4 & 0.85 \\
WD1658+3638 & NOV4 &4 & $11110 \pm 120$&$8.36\pm 0.09$& 19.2 &  0.85\\
WD1726+5331 & NOV7 &1 &  $ 11000 \pm 110$&$ 8.23 \pm 0.08$ & 18.8 & 0.60 \\
\enddata
\end{deluxetable}
\clearpage

\subsection{Uncertainties in temperature and $\log ~g$ determinations}
The true external uncertainties in the SDSS $\teff$ determinations are likely to be larger 
than listed in Paper-I. We expect that the external uncertainties are of the order of 300\,K.
However, the uncertainty that is relevant in determining the width and purity of the instability strip
defined by a homogeneous ensemble is the internal uncertainty.

The low signal-to-noise of the blue end of the SDSS spectra reduces
the reliability of the $\log ~g$ values. The H8 and H9 lines depend mostly on gravity
because neighboring atoms predominantly affect higher energy levels (Hummer \& Mihalas 1970), and their
density depends on $\log ~g$.
The external uncertainties in $\log ~g$ for our ensemble may be as high as
0.1, twice that of the estimated uncertainty for the Bergeron \etal (2004) sample.
We find an average $\log ~g$ of $\simeq 8.10$
for our sample of 31 objects, while the 36 objects in Bergeron \etal
(2004) average at $\simeq 8.11$. G\,238-53 is common to both samples;
Bergeron \etal
(2004) derive ${\mathrm{T}}_{\mathrm{eff}}$=11890\,K and $\log ~g$=7.91, while
the SDSS determination places G\,238-53 at
$\teff = 11820 \pm 50$ and $\log ~g = 8.02 \pm 0.02$. The temperature values agree within $1\sigma $ uncertainties.
Temperature is mainly determined by the continuum and the $H\alpha $, $H\beta $, and $H\gamma $
lines; the low S/N at the blue end of the SDSS spectra
has a reduced effect on temperature determinations.
The well calibrated continuum, extending from 3800--9200\,\AA\, provides an accurate
temperature determination.

The gradual change in mean
mass as a function of temperature for the SDSS DA white dwarf fits is addressed in Kleinman \etal (2004), and
Figure 7 of their paper shows a quantitative measure of this systematic effect.
The increase in $\log ~g$
across the width of the instability strip is only $\sim 0.02$, and implies that our determinations of cDAV
masses are negligibly higher.
These systematic effects are small in the range of the ZZ Ceti
instability strip, and cannot produce
either the narrow width or the impurity of the observed strip.

We conduct a simple Monte Carlo simulation to estimate the internal $\teff$ uncertainties of our ensemble.
Using the observed pulsation characteristics, we can separate the DAVs into two classes:
hDAVs and cDAVs (see section 2.2). We show the observed distribution of the hDAVs
and cDAVs in the lowest panel of Figure 2. These distributions are distinct, except for
3 objects.
Based on the empirical picture, we conceive that the underlying DAV distribution may look similar to
that shown in the topmost panel of Figure 2. We perform a Monte Carlo simulation, drawing
hDAVs and cDAVs randomly from the expected DAV
distribution, and using Gaussian uncertainties with $\sigma = 300$\,K. We show the resulting distribution
in the second panel; the large uncertainties cause significant overlap between the cDAVs and hDAVs, swamping the central
gap. We perform a similar simulation with $\sigma =$200\,K (third panel), and it 
compares well with the observed distribution
considering the small numbers of the empirical distribution. 
This suggests that the internal uncertainties in effective temperature for our ensemble are $\sigma \leq $200\,K
per object,
provided we believe that the hDAVs and cDAVs each span a range of at least 300\,K.
Note that the internal uncertainties for a few individual objects maybe as large as 250--300\,K.

\section{Probing the non-uniform DAV distribution using pulsation periods}
The mean or dominant period of a pulsator is an indicator of its effective
temperature (see section 2.2).
This asteroseismological relation is not highly sensitive, but it provides a technique independent
of spectroscopy to study the DAV temperature distribution.
We show the distribution of the dominant periods of the SDSS DAVs
in Figure 3.
The top right panel in Figure 3 shows the number of DAVs per period interval and is suggestive of a
bimodal distribution; this increases the likelihood that the dearth of DAVs near the center of the strip is real\footnote{We made a similar plot using the dominant periods for the 36 previously known DAVs, 
but did not find any evidence for a bimodal distribution. Determining the dominant period of the
36 \zzc stars in the literature proved to be difficult and quite
inhomogeneous compared to our own data on the SDSS DAVs.}.

\section{Questioning the impurity of the instability strip}
Non-variables in the instability strip imply that all DA white dwarfs do not evolve in the
same way. This notion has a severe implication: decoding the inner structure of a DAV will 
no longer imply that we can use the results towards understanding DA white dwarfs in general.
Hence we question our findings, and conduct simulations to verify our results. 
Although we estimate the internal $\teff$ uncertainties to be at most 200\,K in section 2.3,
we will conservatively assume $\sigma = 300$\,K for all subsequent calculations.

The SDSS spectra do not show any evidence of a binary companion for all the non-variables
within the instability strip.
Also, we used D. Koester's model
atmospheres
to ascertain that the SDSS algorithm had chosen a solution consistent with the photometric colors ($u-g$, $g-r$)
in every case.

We now conduct a Monte Carlo simulation assuming a pure instability strip enclosed by non-variables, as
shown in the top panel of Figure 4. Note that we have not included a $\log ~g$ dependence in our
model, as we expect it to be a smaller effect than what we are about to demonstrate.
We choose non-variables from outside the strip and add uncertainties chosen randomly from a
Gaussian error distribution
with $\sigma =$300\,K to determine the NOV distribution shown in the middle panel.
We find that although non-variables leak into the strip, they are found mostly at the outer edges
and their number tails off within the strip.
The observed NOV distribution (bottom panel) does not show any signs of tailing off
within the instability strip. Rather, it displays the same number of non-variables at the edges as
in the center of the strip. This suggests that the instability strip is impure, and that all
the NOVs within the instability strip did not leak in due to large $\teff$ uncertainties.
We carried out these simulations several times to verify these results.

We compute the
likelihood that the instability strip is pure based on the following two criteria.
There are two ways in which a non-variable can disappear from the instability strip: subsequent observations
show it is a (low amplitude) variable or the internal uncertainties in $\teff$ prove to be large enough to allow the
possibility that it may have leaked into the strip.
Table 1 lists our estimates of the probabilities that the NOVs found within the strip are genuine
non-variables.
The chance that NOVs may have leaked into the strip
due to large internal uncertainties $\sigma =300$\,K are: 
0.35 for WD0037+0031, 0.18 for WD0050-0023, 0.13 for WD0303-0808, 0.04 for WD0345-0036, 0.25 for WD0747+2503,
0.42 for WD0853+0005, 0.15 for WD1031+6122, 0.38 for WD1136-0136, 0.31 for WD1338-0023, 0.11 for WD1342-0159,
0.28 for WD1345+0328, 0.13 for WD1432+0146, 0.25 for WD1503-0052, 0.20 for WD1658+3638, and 0.31 for WD1726+5331.
The probability that each of the above non-variables disappear from the instability strip is then:
0.48, 0.59, 0.26, 0.23, 0.33, 0.68, 0.28, 0.62, 0.41, 0.24, 0.50, 0.30, 0.36, 0.32, and 0.59 respectively.

Three or four of the above non-variables may have an inclination angle that reduces the observed
amplitude below the detection threshold. Instead of calculating various permutations, we will evaluate the likelihood
of the worst
case scenario. Let four NOVs that have the least chance of disappearing from the instability strip be the
ones that have an unsuitable inclination angle for observing pulsations. In that case, the chance that the instability strip is
pure is 0.004\%. 
The impurity of the instability strip suggests that parameters other than
just the effective temperature and $\log ~g$ play a crucial role in deciding the fate
of a DA white dwarf, i.e., whether it will pulsate or not.

\section{Narrow Width of the \zzc strip}
Computing the width of the instability strip using the effective temperatures of the
hottest and coolest pulsators gives us a value, independent of our conception of the shape of the
\zzc strip.
Determining whether the blue and red edges continue to be linear for very
high ($\log ~g\geq 8.5$) or very
low ($\log ~g\leq 7.7$) masses
is presently not possible with either our sample or the Bergeron \etal (2004) sample. The
width of the instability strip calculated from the empirical
edges at different values of $\log ~g$ involves additional uncertainties
from our linear visualization of the edges.

The empirical SDSS DAV instability strip spans from the hottest objects G\,238-53 and WD0825+4119, both at
$\teff=11820 \pm 170$\,K, to the coolest object WD1732+5905 at $10860 \pm 100$\,K.
This span of $960 \pm 200$\,K is considerably smaller than the 1500\,K width in the
literature (Bergeron \etal 1995; Koester \& Allard 2000).
The hottest pulsator in the Bergeron \etal (2004) sample is G\,226-29 at 12460\,K
and the coolest pulsators are G\,30-20 and BPM\,24754 at 11070\,K. The extent of the instability
strip for the Bergeron \etal (2004) sample is then $\sim 1400$\,K.

The drift rates of the stable \zzc pulsators give us a means of measuring their cooling
rates (e.g. Kepler \etal 2000a, Mukadam \etal 2003).
Our present evolutionary cooling rates from such pulsators suggest that given a width of 950\,K,
a 0.6 $M_{\odot }$ \zzc star may spend $\sim 10^{8}$\,yr traversing the instability strip.
This agrees with theoretical calculations by Wood (1995) and Bradley, Winget, \& Wood (1992).
The narrow width
constrains our understanding of the evolution of \zzc stars.

\section{Empirical Blue and Red Edges}
We draw blue and red edges around the DAV distribution that enclose
all of the variables. This is shown in Figure 5 by the solid line for the blue edge
and the line with dots and dashes for the red edge. These edges also include non-variables
within the instability strip.

We now demonstrate an innovative statistical approach
to find the best-fit blue and red edges that maximize the number of variables
and minimize the number of non-variables enclosed within the strip. To the best of our knowledge,
no standard technique can be used to solve this interesting statistical problem.
Our statistical approach has two advantages:
we are accounting for the uncertainties in temperature and $\log ~g$
values and we are utilizing most of the variables and non-variables in our determination rather
than just a handful close to the edge.

This problem has essentially two independent sources of uncertainties: the uncertainties
in temperature and $\log ~g$ that shift the location of a star in the $\teff$--$\log ~g$ plane and the uncertainty concerning the
genuine nature of a non-variable. Pulsators masquerading as non-variables can significantly
alter our determination of the blue and red edges. Hence, we assign different weights
to DAVs and NOVs. 
Since the DAVs are confirmed variables, we assign them a unit weight.
We use the non-variability limit to decide the weight of all the NOVs that lie outside the empirical \zzc 
strip, as in section 2.2,  
while we assign the weights listed
in Table 1 for NOVs within the instability strip.

\subsection{Technique}
We construct a grid in $\teff$ and $\log ~g$ space in the respective ranges 
9000--14000\,K and 6.0--10.5 with resolutions of 50\,K and 0.05.  
For each point in the grid, we consider possible blue and red edges that vary in inclination angle relative
to the temperature axis from 15 degrees to
165 degrees by half a degree with each successive iteration. 

	For each point of the grid, and for each possible blue edge, we compute a net count as follows:
DAVs on the cooler side of the edge count as +1 each and on the hotter side count as $-$1 each.
NOVs on the hotter side of the edge count as $+w$, and on the cooler side as $-w$ each,
where $w$ is the weight
of the corresponding NOV. 
To determine the best
blue edge, we consider
all DAVs and NOVs that satisfy $\teff \geq $11500\,K.
This ensures that the NOVs close to and beyond the red edge do not influence the determination of the blue edge.
If the DAV or NOV is within 3$\sigma $ of the edge, then we determine the net chance that it lies on the hot or cool
side of the edge, assuming a Gaussian uncertainty distribution. We multiply this chance with the count for that object,
before adding it to the total count.
An effect of this choice is that the best edge is determined by the global distribution of DAVs and NOVs,
rather than the few close to the edge.

	Similarly, we determine the best red edge at each point of the grid by counting DAVs
on the hotter side of the edge as +1 and NOVs on its cooler side as $+w$, and vice versa.
We consider all
DAVs and NOVs within the instability strip and cooler than 11820\,K to
compute the best red edge.
If the DAV or NOV is within 3$\sigma$ of
the red edge, then its contribution is a fraction of the above, depending on the probability that it lies
on one side of the edge or the other.

To test our statistical approach, we input the $\teff$ and $\log ~g$ determinations of the previously known
DAVs from Bergeron \etal (2004) along with the SDSS NOVs. The resulting red and blue edges are fairly similar to
those of Bergeron \etal (2004), and we attribute most of the difference to using an independent set of NOVs\footnote{We cannot use
the same set of non-variables as Bergeron \etal (2004) as they did not publish the non-variable parameters
or identifications.}.
Figure 5 shows our best-fit for the red edge and our constraint on the blue edge using our statistical approach.\newline
For the blue edge, we determine:\newline
Best-fit $\log ~g$=4.33 $\teff$ -434.77 \newline
+1$\sigma $ $\log ~g$=1.57 $\teff$ -106.39 \newline
-1$\sigma $ $\log ~g$=3.73 $\teff$ -363.45 \newline

For the red edge, we determine:\newline
Best-fit $\log ~g$=1.036 $\teff$ -30.12 \newline
1$\sigma$ $\log ~g$=1.192 $\teff$ -47.26 \newline

\subsection{Estimating the Uncertainties}
The dominant effect that dictates the uncertainties in the slope ($\log ~g$ dependence)
and location (in temperature) of the edges
arises as a result of the unreliable nature of the NOVs. Are they genuine NOVs or low amplitude pulsators? 
Our simulations in section 2.2 show that we miss 30\% of high mass pulsators due to their low
amplitude. We estimate this should introduce an uncertainty of order 0.2 in the total count for both the red and blue edges.
The NOVs close to the blue edge, but within the instability strip, can introduce additional uncertainties
in our determination.
We add these independent sources of uncertainty in quadrature to obtain an estimated 1$\sigma $ uncertainty of 0.6
for the red edge and 0.4 for the blue edge. We show these as dotted lines in Figure 5.
Our estimates of the 1$\sigma $ uncertainties clearly show that the red edge is well constrained, and the slope
of the blue edge is not.

Note that we already account for the uncertainties in $\teff$ and $\log ~g$ in determining the red and blue edges.
The unreliability of these uncertainties contributes towards an uncertainty in the slope of the edges; this turns
out to be a negligible second order effect.

\subsection{Comparison with Empirical Edges}
We show the empirical blue and red edges from Bergeron
\etal (2004) in Figure 5 for comparison. The slopes of the red edges from both samples agree within
the uncertainties. But our constraint on the blue edge differs significantly from that of Bergeron \etal (2004), and
suggests that the dependence on mass is less severe.

The mean temperature of our sample is 11400\,K, while the mean temperature for the
Bergeron \etal (2004) sample is 11630\,K. The observed extent of our instability strip defined
by 31 objects spans 10850--11800\,K, while that of Bergeron \etal (2004) spans 11070--12460\,K\footnote{Excluding
G\,226-29, the Bergeron \etal (2004) sample spans a width of 1060\,K from 11070\,K to 12130\,K.}.
We can consider these values to imply a relative shift of $\sim $200\,K between our sample and that of Bergeron \etal (2004).

We would also like to point out that our sample is magnitude limited and reaches out to
$g=19.3$. We are effectively sampling a different population of stars, more distant by a factor of 10, than the Bergeron
\etal (2004) sample.

\subsection{Comparison with Theoretical Edges}
In Figure 5, we show the theoretical blue edge from Brassard \& Fontaine
(1997) due to the traditional radiative driving mechanism; they
use a ML2/$\alpha=$0.6 prescription for convection in their equilibrium
models. We also show the blue \emph{and} red edges which we derive
from the convective driving theory of Wu \& Goldreich (Brickhill 1991; Wu 1998;
Wu \& Goldreich 1999), assuming ML2/$\alpha=$0.8
for convection.

We see that the blue edges of the two theories are essentially
the same, and would nearly coincide if the mixing-length parameter
were tuned. To obtain the red edge of Wu \& Goldreich, we have made
the following assumptions: (1) the relative flux variation at the base
of the convection zone is no larger than 50\%, (2) the period of a
representative red edge mode is 1000 s, and (3) the detection limit
for intensity variations is 1\,mma. Within this theory,
the convection zone attenuates the flux at its base by a factor of
$\sim \omega \tau_C$, where $\tau_C$ is the thermal response time of
the convection zone, so we have adjusted $\tau_C$ such that the surface
amplitude $0.5/(\omega \tau_C) \sim 10^{-3}$, equal to the detection
threshold.

	The observed distribution of variables and non-variables suggests that
the mass dependence of the blue edge is less severe than predicted by the models.
Both the slope and the location of the red edge we calculate
are consistent with the observed variables and non-variables
within the uncertainties.

\section{Conclusion}
Using a statistically significant and truly homogeneous set of 31 \zzc spectra, we find a narrow
instability strip between 10850\,K and 11800\,K. We also find non-variables within the strip and compute
the likelihood that the instability strip is pure to be $\sim 0.004$\%. 
Obtaining higher signal-to-noise spectra of all the SDSS and non-SDSS DAVs as well as non-variables
in the \zzc strip is crucial to improving our determination of
the width and edges of the instability strip, and in investigating the purity of the
instability strip. This should help constrain our understanding of
pulsations in \zzc stars. 

The DAV distribution shows a scarcity 
of DAVs in the range 11350--11500\,K. After exploring various possible causes for such a bimodal, non-uniform 
distribution, we are still not entirely confident that it is real. The data at hand are suggestive
that the non-uniformity of the DAV distribution is real, and stayed hidden from us for decades due to the
inhomogeneity of the spectra of the previously known DAVs.
However, we are in the domain of small number
statistics and unless we investigate additional targets in the middle of the strip, we cannot
be confident that the bimodal distribution is not an artifact in our data.

\acknowledgements
We thank R. E. Nather for useful discussions that benefited this paper. We also thank
J. Liebert and the referee in helping us improve our presentation considerably.
We thank the Texas Advanced Research Program for the grant ARP-0543,
and NASA for the grant NAG5-13094 for funding this project.
We also thank the UT-CAPES international collaboration for their funding and support.

{}

\onecolumn

\clearpage
\begin{figure}[h]
\figurenum{1}
\psfig{figure=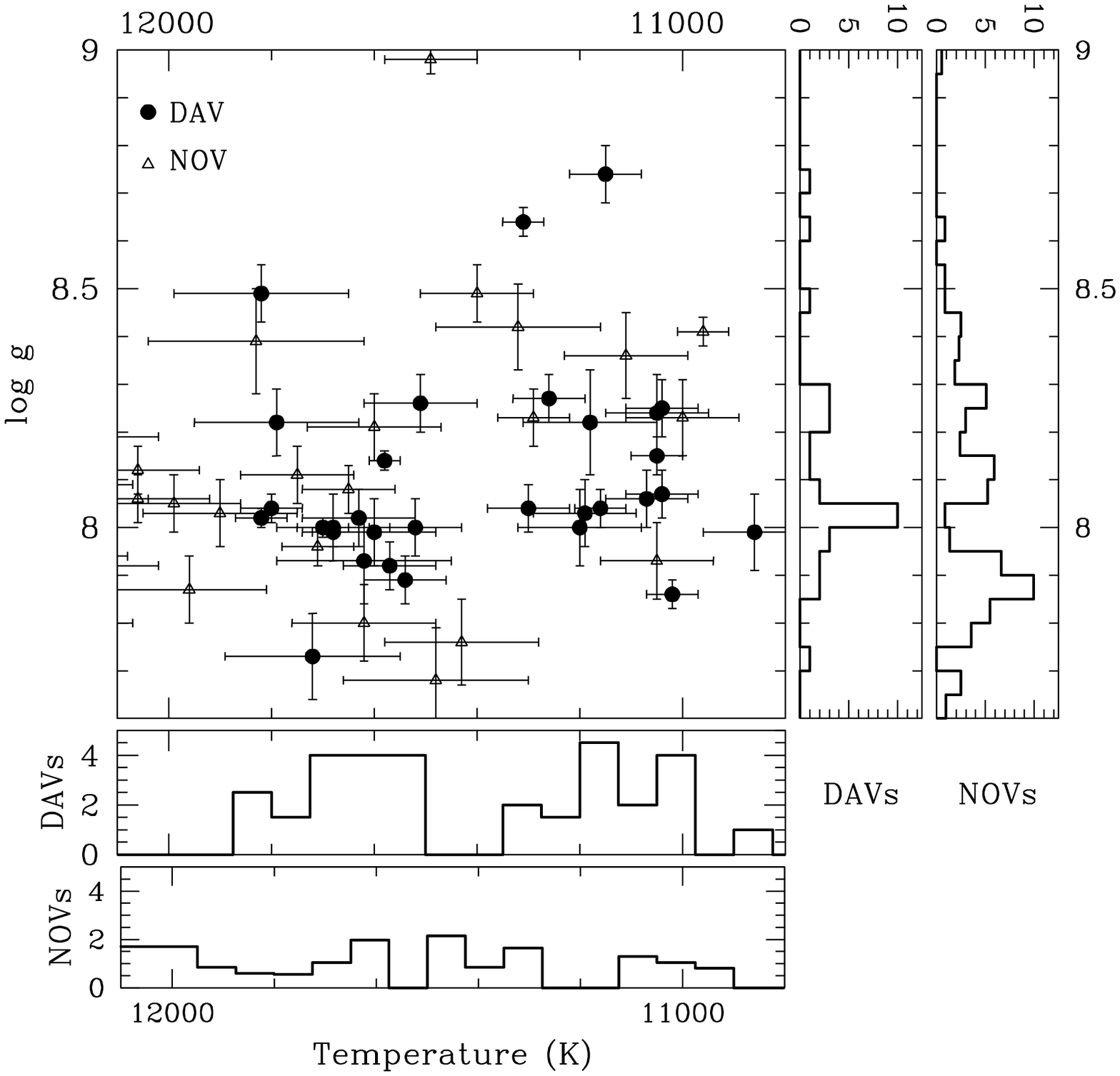,width=6.5in}
\figcaption{The distribution of new SDSS DAVs and NOVs (Mukadam \etal 2004) as
a function of temperature and $\log ~g$. We also include G\,238-53 in this plot.
The narrow width of the instability strip and the presence of non-variables within
form the two prominent features of this figure. We also note the
paucity of DAVs in the middle of the instability strip.}
\end{figure}

\begin{figure}[h]
\figurenum{2}
\psfig{figure=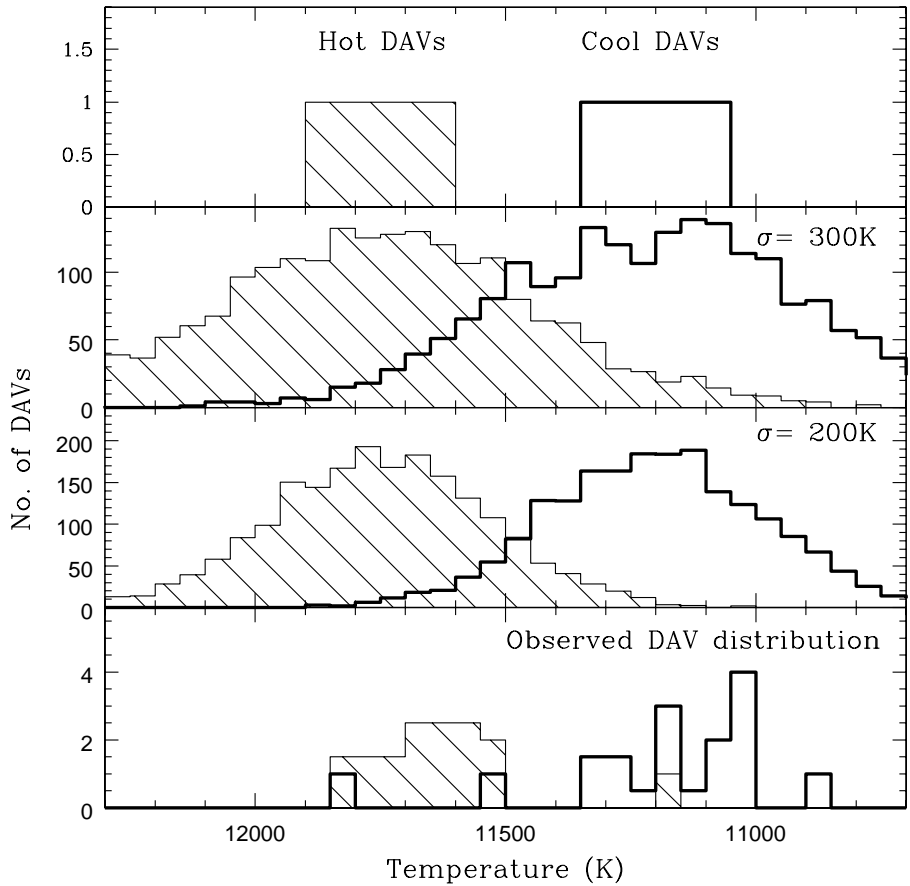,width=5.0in}
\figcaption{We choose hDAVs and cDAVs from the distributions
shown in the top panel, and use a Gaussian error function with
$\sigma = 300$\,K to compute the distributions shown in the
second panel. We also similarly determine a DAV distribution with internal
uncertainties of order 200\,K, shown in the third panel. Comparing
the empirical DAV distribution, shown in the bottom panel, to the
synthetic computations, we conclude that the average internal uncertainty
for our ensemble is $\sigma \leq 200$\,K.}
\end{figure}

\begin{figure}[h]
\figurenum{3}
\psfig{figure=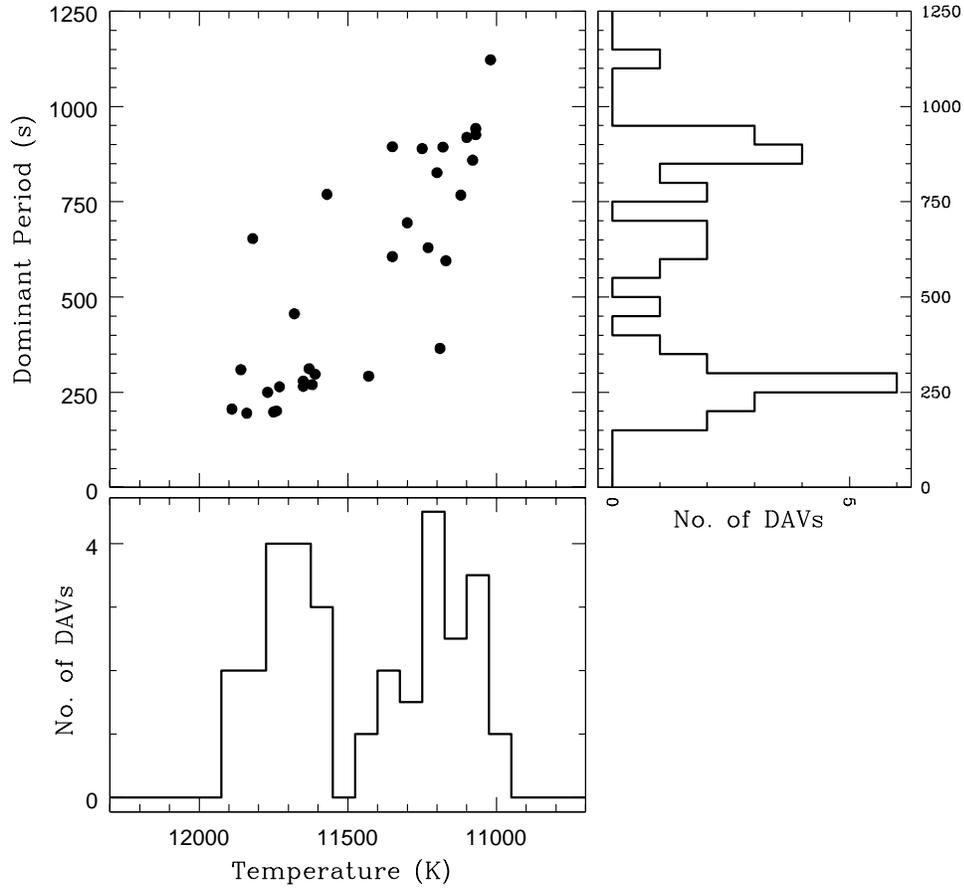,width=5.0in}
\figcaption{Period distribution of the SDSS DAVs as a function of temperature. The top
left panel exhibits two distinct clumps consisting of the short period hDAVs and the long period cDAVs. The
dominant period of a DAV is a seismological temperature indicator and the histogram shown in the top
right panel is suggestive of a bimodal distribution.}
\end{figure}

\begin{figure}[h]
\figurenum{4}
\psfig{figure=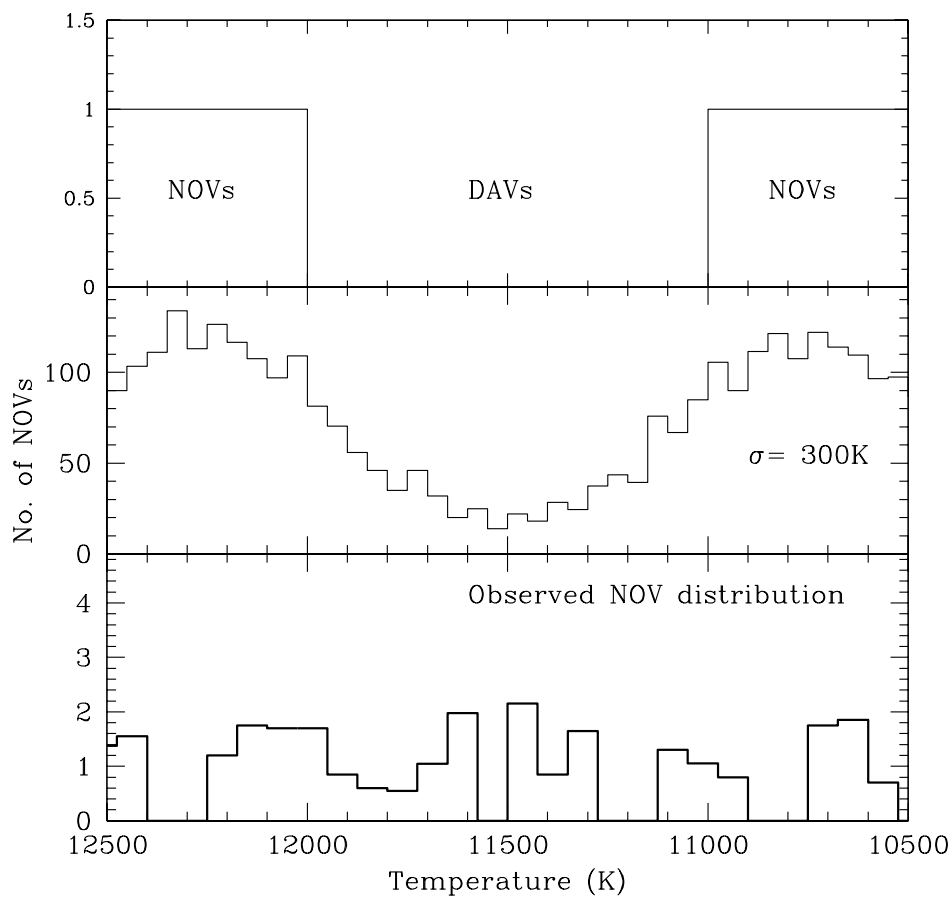,width=5.0in}
\figcaption{Assuming a pure instability strip as shown in the top panel, we use a Monte Carlo simulation
assuming a Gaussian distribution for the internal uncertainties with $\sigma =$300\,K 
to determine the expected distribution for non-variables within the strip.
The observed NOV distribution is flat, and shows no signs of tailing off within the
strip. The observed distribution shows the same number of non-variables at the edges as
in the center of the instability strip.}
\end{figure}

\begin{figure}[h]
\figurenum{5}
\psfig{figure=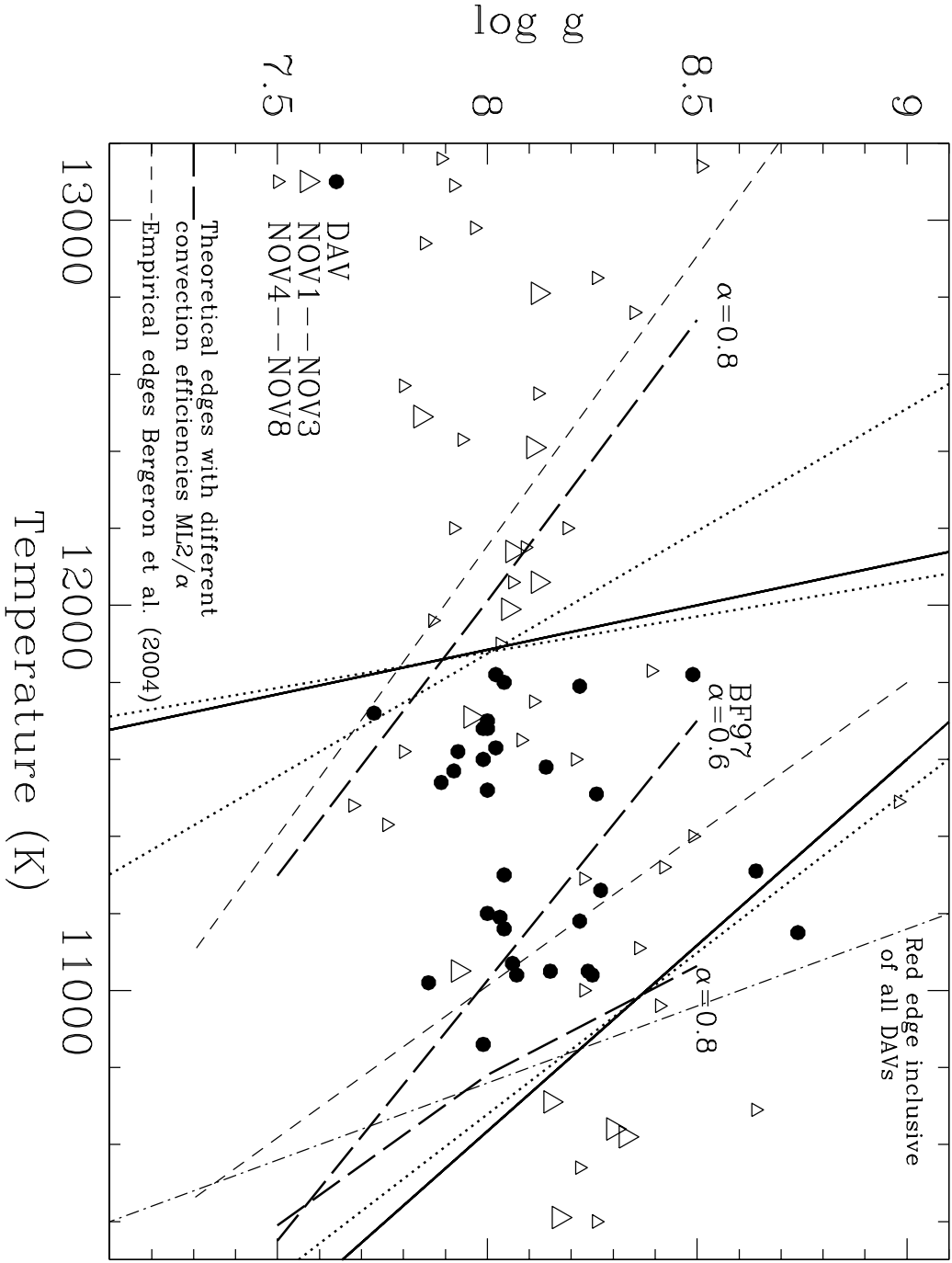,width=5.7in,angle=180}
\figcaption{Statistical determination of the blue and red edges from the homogeneous
set of 31 SDSS DAVs. The thick solid line shows the global solution, while the two dotted lines on
either side show
the estimated $1\sigma $ uncertainty in our
determination. Note that the red edge is coincident with one of the dotted lines.
Although our blue edge does not exclude any DAVs, our best-fit red edge does.
We present the line shown on the extreme right with dots and dashes as a red edge inclusive of all DAVs.
We also show the empirical blue and red edges from Bergeron \etal (2004) as dashed lines, and
the theoretical blue edge from Brassard \& Fontaine (1997; ML2/$\alpha$=0.6) for
comparison.
We show our computations of the theoretical blue and red edges assuming
ML2/$\alpha$=0.8 convection, based on the convective driving theory of Wu \& Goldreich (1999).}
\end{figure}

\end{document}